# Hawking-Unruh effect and the entanglement of two-mode squeezed states in Riemannian spacetime

D. Ahn* & M. S. Kim †

*Institute of Quantum Information Processing and Systems, University of Seoul, Seoul 130-743, Republic of Korea*

†*School of Mathematics and Physics, The Queen's University, Belfast BT7 1NN, United Kingdom*

*Abstract:* We consider the system of free scalar field, which is assumed to be a two-mode squeezed state from an inertial point of view. This setting allows the use of entanglement measure for continuous variables, which can be applied to discuss free and bound entanglement from the point of view from non-inertial observer.

*Corresponding author:*

Doyeol (David) Ahn, Ph. D., Professor, IEEE Fellow
Department of Electrical and Computer Engineering
University of Seoul
Seoul 130-743, Republic of Korea
Tel: int.+82-2-2210-2468(office), Mobile: int.+82-(0)11-434-4531
Fax: int. +82-2-2210-2692, e-mail:dahn@uos.ac.kr ; davidahn@hitel.net

The Hawking-Unruh effect [1-4], a quantum thermal radiation from a black hole, is the origin of information paradox [5-10] that has been one of the major problems in theoretical physics. The key question is whether a pure quantum state decays into a mixed state or there is any correlation or quantum information left as the quantum system falls into the event horizon. The basic entity of quantum information is the entanglement [11]. The quantitative measure of quantum entanglement is the entanglement monotone [12-14], which describes the degree of entanglement.

The setting, in which Alice and Rob are two observers, one inertial and the other non-inertial, who describe the entanglement between two modes of free scalar field from the point of view of their detectors, is especially important because it is directly related to the black hole information paradox. The issue is that black holes appear to absorb the quantum information as well as the matter, yet the most fundamental laws of physics demand that this information should be preserved as the universe evolves. When a non-inertial observer is under the influence of strong gravitational field near the event horizon of the black hole, the measure of entanglement seen by non-inertial observer is affected by the presence of quantum thermal fields known as the Hawking-Unruh effect. The essential feature of the Hawking-Unruh effect apart from the complications due to the curvature of the spacetime of the black hole is contained in the much simpler situation involving the uniform acceleration of Rob in the flat spacetime, so called Rindler spacetime [15-17]. The Schwarzschild spacetime near the event horizon resembles Rindler spacetime in the infinite acceleration limit. A consequence of Hawking-Unruh effect is that an entangled pure state seen by inertial observers appears mixed from an accelerated frame. There have been studies on the behaviour of entangled qubits coupled to the Hawking-Unruh field in the Rindler spacetime [18, 19]. They considered the entanglement between two modes of free scalar field when one of the observers describing the state is uniformly accelerated. The state observed by an inertial observer Alice and a non-inertial observer Rob is in $2 \times \infty$ dimensional space in which case we do not have the necessary and sufficient criteria for the entanglement [20]. That was the reason why the recent studies could give only the lower bounds of an entanglement.

In this paper, we consider the system of free scalar field, which is assumed to be a two-mode squeezed state from an inertial point of view. Non-inertial observer Rob would always detect a Gaussian state for which we have a well-defined entanglement condition and measure of entanglement. When a quantum system is coupled to Hawking radiation, it is inevitably treated in an infinite dimensional space, in which case only a Gaussian state has an entanglement measure. Even though their prediction was relatively correct, Alising and Milburn [18] used only an indirect measure of entanglement as they calculated a teleportation fidelity. Fuentes-Schuller and Mann [19] also calculated lower bound of entanglement. The entanglement measured by Rob is decreasing as Rob is uniformly accelerating and is approaching to zero in the infinite acceleration limit.

***Hawking-Unruh effect in Rindler spacetime.*** We consider the real, scalar field both in Minkowski and Rindler spacetime. Let Alice be an observer at event *P* with zero velocity in Minkowski spacetime and non-inertial observer Rob be moving with positive uniform acceleration in the z direction with respect to Alice (Fig. 1). If Rob is under a uniform acceleration, the corresponding ground state should be specified in Rindler coordinate [15-17] in order to describe what Rob observes. Let us denote the ground states, which Alice and Rob detect in Minkowski spacetime as $|O_A\rangle_M$ and $|O_R\rangle_M$ (Fig. 1), respectively. Then ground state from the non-inertial point of view can be written as $|O_R\rangle_M = \frac{1}{\cosh r}\sum_{n=0}^{\infty}\tanh^n r |n\rangle_I \otimes |n\rangle_{II}$, with *r* the acceleration parameter defined by $\tanh r = \exp(-2\pi\Omega)$, $\Omega = |k|c/a$, *k* the wave vector, *c* the speed of light, *a* the uniform acceleration, and $|n\rangle_I$ and $|n\rangle_{II}$ the mode decompositions in Rindler regions *I* and *II*, respectively. The excited state for Rob in Minkowski spacetime is obtained by applying the Minkowski creation operator $a_R^\dagger$ to the vacuum state successively. For example,

$$|1_R\rangle_M = a_R^\dagger |O_R\rangle_M,\ |2_R\rangle_M = \tfrac{1}{\sqrt{2!}}(a_R^\dagger)^2|O_R\rangle_M,\ \ldots\ |m_R\rangle_M = \frac{1}{\sqrt{m!}}(a_R^\dagger)^m|O_R\rangle_M. \qquad (1)$$

The particle creation and annihilation operators for the Rindler space-time are expressed as $b_\sigma^\dagger$ and $b_\sigma$, respectively. Here, the subscript $\sigma = I$ or $II$, takes into account the fact that the space-time has an event horizon, so that it is divided into two causally disconnected Rindler wedges *I* and *II* (Fig. 1). The Minkowski operators $a_R^\dagger$ and $a_R$ can be expressed in terms of the Rindler operators $b_\sigma^\dagger$ and $b_\sigma$ by Bogoliubov transformations [15-17]:

$$a_R^\dagger = b_I^\dagger \cosh r - b_{II}\sinh r = Gb_I^\dagger G^\dagger,\quad a_R = b_I\cosh r - b_{II}^\dagger \sinh r = Gb_I G^\dagger, \qquad (2)$$

with $G = \exp\{r(b_I^\dagger b_{II}^\dagger - b_I b_{II})\}$. Then, the Minkowski ground state $|O_R\rangle_M$ seen by the Rindler observer, i.e., Rob, is given by $|O_R\rangle_M = G(|O\rangle_I \otimes |O\rangle_{II})$. This is the basis of the Hawking-Unruh effect [1-4], which says that a non-inertial observer with uniform acceleration (or an observer near the event horizon of the black hole) would see thermal quantum fields. In other words, Rob would see the quantum bath populated by thermally excited states. The excited states for Rob in Minkowski spacetime are now given by

$$a_R^\dagger |O_R\rangle_M = G b_I^\dagger (|O\rangle_I \otimes |O\rangle_{II}) \ldots, (a_R^\dagger)^m |O_R\rangle_M = G(b_I^\dagger)^m (|O\rangle_I \otimes |O\rangle_{II}). \quad (3)$$

***Hawking-Unruh effect on a two-mode squeezed state.*** We consider the system of a free scalar field, which is assumed to be in a two-mode squeezed state [21] with squeezing parameter $s$, from an inertial point of view:

$$|\Psi\rangle = \frac{1}{\cosh s} \sum_{m=0}^{\infty} \tanh^m s \frac{1}{m!} (a_A^\dagger a_R^\dagger)^m (|O_A\rangle_M \otimes |O_R\rangle_M)$$
$$= S_{AR}(s)(|O_A\rangle_M \otimes |O_R\rangle_M) \quad (4)$$

with $S_{AR}(s)$ defined by $S_{AR}(s) = \exp\{s(a_A^\dagger a_R^\dagger - a_A a_R)\}$.

Rob's trajectory is a hyperbola in the right Rindler wedge labelled region $I$, bounded by the asymptotes $H_-$ and $H_+$, which represent Rob's past and future horizons (Fig. 1). The state Rob observes must be restricted to the right Rindler wedge, i.e., region I, in which his motion is confined. The two-mode squeezed state $|\Psi\rangle$ seen by Rob is now described by $|\Psi\rangle = G S_{AI}(s)(|O_A\rangle_M \otimes |O\rangle_I \otimes |O\rangle_{II})$. If we interpret $G$ as a unitary operator corresponding to the uniform acceleration of non-inertial observer Rob and $S_{AI}(s)$ as a squeezing operator for Alice in Minkowski spacetime (and Rob confined to region $I$), the state $|\Psi\rangle$ is the result of squeezing the state of Alice and Rob (in region $I$). It is now obvious that Rob always detects Gaussian states. In this case, we have a well-defined and analytical measure of entanglement.

***Entanglement measure.*** In order to find the entanglement condition, we need to find the variance matrix $V$ whose element $V_{ij}$ is defined by [14]

$$V_{ij} = \frac{1}{2}\langle\{X_i, X_j\}\rangle = \frac{1}{2}\langle X_i X_j + X_j X_i\rangle, \quad (5)$$

where vector $X = (q_A, p_A, q_I, p_I)$, with the operator defined as $a_A = (q_A + ip_A)/\sqrt{2}$ and $b_I = (q_I + ip_I)/\sqrt{2}$. For the calculation, we also need to know the following transformations [21]:

$$S_{AI}^\dagger G^\dagger q_A G S_{AI} = q_A \cosh s - q_I \sinh s,$$
$$S_{AI}^\dagger G^\dagger p_A G S_{AI} = p_A \cosh s + p_I \sinh s,$$
$$S_{AI}^\dagger G^\dagger q_I G S_{AI} = (q_I \cosh s - q_A \sinh s)\cosh r - q_{II} \sinh r,$$
$$S_{AI}^\dagger G^\dagger p_I G S_{AI} = (p_I \cosh s + p_A \sinh s)\cosh r + p_{II} \sinh r.$$

As an example, we calculate $V_{11} = \langle 000|S_{AI}^\dagger G^\dagger q_A^2 G S_{AI}|000\rangle = \frac{1}{2}\cosh 2s$, with $|000\rangle \equiv |O_A\rangle_M \otimes |O\rangle_I \otimes |O\rangle_{II}$. After a straightforward calculation, the other elements can also be found and the final variance matrix is given by

$$V = \frac{1}{2}\begin{pmatrix} A & 0 & -C & 0 \\ 0 & A & 0 & C \\ -C & 0 & B & 0 \\ 0 & C & 0 & B \end{pmatrix}, \qquad (6)$$

where $A = \frac{1}{2}\cosh 2s$, $B = \cosh^2 r \cosh^2 s - \frac{1}{2}$, and $C = \frac{1}{2}\sinh 2s \cosh r$.

Let us define the symplectic eigenvalue [14] $\tilde{\lambda}$ of the matrix obtained from V through the partial transposition (PT): $\tilde{\lambda} = \frac{1}{\sqrt{2}}[\Sigma - (\Sigma^2 - 4\det V)^{1/2}]^{1/2}$ where $\Sigma$ is given by $A^2 + B^2 + 2C^2$. It can be shown that the minimum PT symplectic eigenvalue represents an entanglement monotone, $E_N$ which describes the degree of entanglement and is given by $E_N = \max[0, -\ln 2\tilde{\lambda}]$. The Gaussian state is entangled if and only if $\tilde{\lambda} < 1/2$, which is equivalent to $E_N > 0$ [14].

In Fig. 2, we plot the entanglement measure $E_N$ as a function of the acceleration $r$ for three different initial values of the squeezing parameter. The monotonous decrease of $E_N$ with increasing $r$ indicates that quantum coherence of the initial squeezed state is lost to the thermal fields generated by the Hawking-Unruh effect. In essence, the Hawking-Unruh fields act as heat baths for the initial squeezed state. It is also interesting to note that the stronger the initial squeezing, the faster it loses quantum coherence. In the asymptotic limit $r \to \infty$, we obtain

$$2(\tilde{\lambda})^2 \to \frac{\cosh^2 2s}{2} + \frac{\sinh^2 2s}{2\cosh^2 s}\left(-\cosh 2s + \frac{\sinh^2 2s}{4\cosh^2 s}\right) = \frac{1}{2}, \qquad (7)$$

and $E_N = -\ln 2\tilde{\lambda}^- \to 0$.

It is interesting to note that entanglement completely disappears in this senario. This should be compared with the paper by Fuentes-Schuller and Mann [19] as they could not definitely conclude this because they could not discuss bound entanglement. However, our clear conclusion is due to the fact that there is no bound entanglement for a Gaussian state. In Eq. (6), *C* being non-zero means the states detected by Alice and Rob are not independent. Even in the infinite acceleration limit, there remains some classical correlation between the states detected by Alice and Rob.

We studied whether there is any entanglement between Alice and the state in Rindler region *II*, denoted as Rob (*II*) in the following, by calculating $V'_{ij} = \frac{1}{2}\langle\{Y_i, Y_j\}\rangle$ with the vector $Y = (q_A, p_A, q_{II}, p_{II})$. Here, we found $V'_{13} = V'_{24} = \frac{1}{2}\sinh r \sinh 2s$, which indicates that Alice is never entangled with Rob (*II*). One of the important conditions for Gaussian entanglement is that the sign of $V'_{13} \neq$ the sign of $V'_{24}$. We also studied the entanglement of Rob (region *I*), denoted as Rob (*I*) with Rob (*II*) by calculating $V''_{ij} = \frac{1}{2}\langle\{Z_i, Z_j\}\rangle$ with $Z = (q_I, p_I, q_{II}, p_{II})$. It is straightforward to show that $V''_{13} = -V''_{24} = -\frac{1}{2}\sinh 2r \cosh^2 s$. The entanglement of Rob (*I*) with Rob (*II*) grows linearly with respect to *r*. It also depends on the initial squeezing parameterized by *s*.

The results indicate that the entanglement between Alice and Rob (*I*) is not lost due to an entanglement between Alice-Rob (*II*) but due to entanglement between Rob (*I*) and Rob (*II*). This process somehow reduces the entanglement but Alice-Rob (*I*)-Rob (*II*) entanglement is still there while the bipartite entanglement between Alice and Rob (*I*) disappears. Quantum coherence is destroyed by the quantum thermal field, i.e., the Hawking-Unruh effect.

The main findings in this paper are due to our choice of the initial Gaussian entangled state: 1) in contrast to the earlier works, by starting with Gaussian entangled states, we have a sufficient and necessary condition for entanglement which has enabled us to see clearly that there is no entanglement when the observer's acceleration is infinity. 2) With the same reason, we have a well-defined measure of entanglement with which we have found that when the initial entanglement is stronger, we lose it more rapidly. Another new result presented in this paper is that entanglement is degraded to a higher degree when entanglement in the inertial frame is higher.

**Acknowledgements** This work was supported by the Korea Science and Engineering Foundation, the Korean Ministry of Science and Technology through the Creative Research Initiatives Program.

**Figure captions**

**Figure 1** Rindler spacetime. In region *I* and *II*, time coordinates $\eta =$ constant are straight lines through the origin. Space coordinates $\zeta =$ constant are hyperbolae with null asymptotes $H_+$ and $H_-$, which act as event horizons. The Minkowski coordinates *t*, *z* and Rindler coordinates $\eta, \zeta$ are given by $t = a^{-1}\exp(a\zeta)\sinh a\eta$ and $z = a^{-1}\exp(a\zeta)\cosh a\eta$, where *a* is a uniform acceleration (Reference 21). We assume that Alice is stationary and Rob (green hyperbola) is under uniform acceleration.

**Figure 2** Measure of entanglement versus acceleration r. The measure of entanglement $E_N = \max[0, -\ln 2\tilde{\lambda}^-]$ is calculated as a function of the acceleration *r* with different initial squeezing parameter *s*. In the limit $r \to \infty$, the results agree well with the asymptotic form given by equation (7).

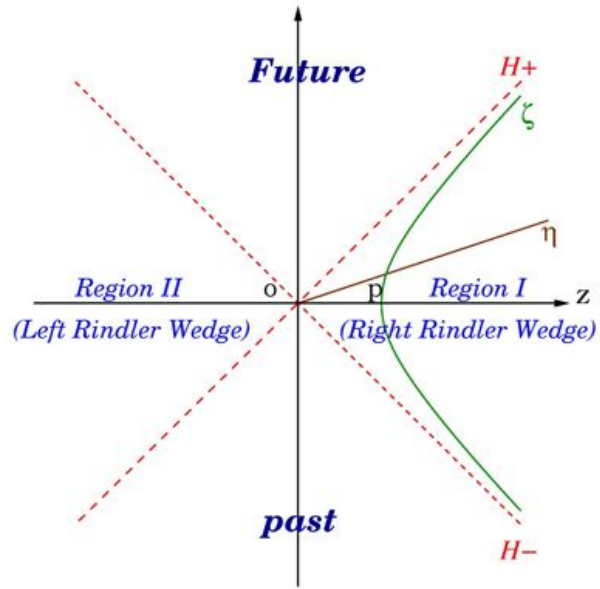

Fig. 1

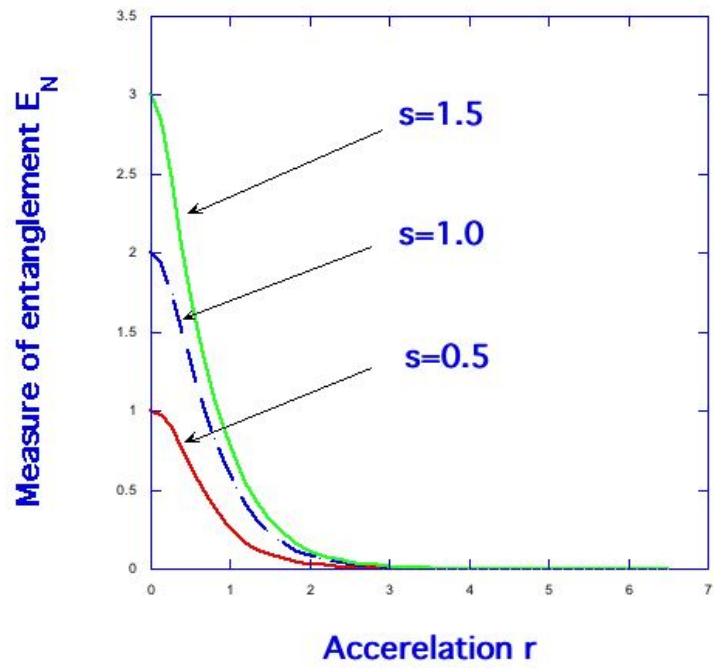

Fig. 2